\documentclass[aps,pra,twocolumn,showpacs,superscriptaddress,nofootinbib,10pt]{revtex4-1}

\usepackage{xcolor}
\usepackage{braket}
\usepackage{slashed}
\usepackage{graphicx}
\usepackage{mathtools}
\usepackage{nicefrac}
\usepackage{amssymb}
\usepackage{mathrsfs}
\usepackage{IEEEtrantools}
\definecolor{BrickRed}{cmyk}{0, .89, .94, .28} 

\newcommand{\etal}{\textit{et al. }}
\newcommand{\etalcomma}{\textit{et al., }}

\newcommand{\be}{\begin{equation}}
\newcommand{\ee}{\end{equation}}
\newcommand{\bea}{\begin{IEEEeqnarray}{rCl}}
\newcommand{\eea}{\end{IEEEeqnarray}}

\newcommand{\ba}{\begin{array}}
\newcommand{\ea}{\end{array}}

\newcommand{\G}{\mathcal{G}(z)}
\newcommand{\g}{\mathcal{G}_0(z)}
\newcommand{\gu}{\mathcal{G}_u(z)}

\definecolor{lightGray}{RGB}{220,220,220}

\definecolor{MyDarkBlue}{rgb}{0,0.08,0.45}

\definecolor{MyDarkGray}{RGB}{140,140,140}

\newcommand{\bes}{\begin{equation}\begin{split}}
\newcommand{\ees}{\end{split}\end{equation}}

\definecolor{lightGray}{RGB}{220,220,220}
\definecolor{MyDarkGray}{RGB}{140,140,140}

\definecolor{MyDarkBlue}{rgb}{0,0.08,0.45}

\newlength{\eqboxstorage}

\newcommand{\ed}{\end{document}}

\newcommand{\bl}{\begin{list}{$\circ$}{}}
\newcommand{\el}{\end{list}}

\definecolor{cadmiumgreen}{rgb}{0.0, 0.42, 0.24}

\usepackage{graphicx}
\usepackage{dcolumn}
\usepackage{bm}
\usepackage{xcolor}

\newcommand{\GF}{Green's function}
\definecolor{newRed}{rgb}{0.6, 0.088, 0.176}

\usepackage{enumitem}
\usepackage{tikz}
\usetikzlibrary{decorations.markings}

\usepackage{hyperref}

\begin{document}

\title{The self-consistent Dyson equation and self-energy functionals: failure or
new opportunities?}

\author{Walter Tarantino}
\email{walter.tarantino@polytechnique.edu}
\affiliation{Laboratoire des Solides Irradi\'es, Ecole Polytechnique, CNRS, CEA/DSM
and European Theoretical Spectroscopy Facility (ETSF), 91128 Palaiseau, France.}
\author{Pina Romaniello}
\affiliation{Laboratoire de Physique Th\'eorique, CNRS, IRSAMC, Universit\'e
Toulouse III - Paul Sabatier and European Theoretical Spectroscopy Facility
(ETSF), 118 Route de Narbonne, F-31062 Toulouse Cedex, France}
\author{J. A. Berger}
\affiliation{Laboratoire de Chimie et Physique Quantiques, IRSAMC, Universit\'e
Toulouse III - Paul Sabatier, CNRS and European Theoretical Spectroscopy Facility
(ETSF), 118 Route de Narbonne, F-31062 Toulouse Cedex, France}
\author{Lucia Reining}
\affiliation{Laboratoire des Solides Irradi\'es, Ecole Polytechnique, CNRS, CEA/DSM
and European Theoretical Spectroscopy Facility (ETSF), 91128 Palaiseau, France.}

\date{\today} 

\begin{abstract}
Perturbation theory using self-consistent Green's functions is one of the
most widely used approaches to study many-body effects in condensed
matter. On the basis of general considerations and by performing
analytical calculations for the specific example of the Hubbard atom, we
discuss some key features of this approach. We show that when the domain of
the functionals that are used to realize the map between the
non-interacting and the interacting Green's functions is properly defined,
there exists a class of self-energy functionals for which the
self-consistent Dyson equation has only one solution, which is the
physical one. We also show that manipulation of the perturbative
expansion of the interacting Green's function may lead to a wrong
self-energy as functional of the interacting Green's function,
at least for some regions of the parameter space.
These findings confirm and explain numerical results of Kozik \etal
for the widely used skeleton series of Luttinger and
Ward [Phys. Rev. Lett. \textbf{114}, 156402]. Our study shows that it is important to
distinguish between the maps between sets of functions and the functionals
that realize those maps. We demonstrate that the self-consistent
Green's functions approach itself is not problematic, whereas the
functionals that are widely used may have a limited range of validity.
\end{abstract}

\maketitle

\section*{Introduction}

In quantum many-body theory observables can in principle be calculated 
as expectation values or weighted sums over expectation values of operators,
using many-body wavefunctions. However, a many-body wavefunction for more than 
a few electrons is a huge object, much too large to be calculated 
or even stored \cite{kohn1999}. Therefore several ways
have been developed to avoid the explicit calculation of wavefunctions. 
One of the strategies is the use of \textit{functionals}: 
it relies on the insight that observables can in principle be expressed 
as functionals of some quantities $Q$ that are more compact than the full 
$N$-body wavefunction $\Psi({\bf r}_1,...{\bf r}_N)$. 
The problem is then split in three parts: first, to find the quantity $Q$ that 
encodes sufficient information to calculate the wanted observable $O$, then
to work out the functional relation between $Q$ and $O$, $O=F[Q]$, 
and, finally, to determine the value $\tilde{Q}$ that a system takes, 
such that $\tilde{O}=F[\tilde{Q}]$ can be calculated.
The most striking example is density functional theory (DFT),
where the charge density $n({\bf r})$ plays the role of $Q$.
Although $n$ has much less degrees of freedom than  $\Psi({\bf r}_1,...{\bf r}_N)$, the
Hohenberg-Kohn theorem guarantees that it encodes all information
necessary for calculating any wanted observable. 
Moreover, the Kohn-Sham approach provides a handy tool to get
a sufficiently accurate estimate of $n({\bf r})$ for any system.
However, the explicit expression of almost all observables in terms of $n({\bf r})$ is unknown. 
An alternative to $n({\bf r})$ is the equilibrium one-body \GF\
$G({\bf r}, \sigma, t, {\bf r}', \sigma', t')$. Contrary to the density, 
it is non-local in space, spin and time (or frequency dependent, for its Fourier transform), 
but it is still much more compact than the many-body wavefunction. 
Moreover, it has the advantage that all expectation values of one-body operators, 
as well as the total energy, have a simple, known expression in terms of it.

To determine the value that the one-body \GF\ takes for the system of interest,
a common approach is to solve the Dyson equation,\footnote{%
	For emphasizing the algebraic structure of the equations,
    spin indices, space and time coordinates, as well as integrals and summations,
    and exact numerical coefficients will be omitted henceforth in this section. 
    This will also apply to equations (\ref{Gper}-\ref{scdex}) and (\ref{Gper2}-\ref{Sigmaper}) of later sections.}
    $G=G_0 + G_0 \Sigma G$,
where $\Sigma$ is a non-local and frequency-dependent function, called self-energy. 
Perturbation Theory offers a way to
calculate the (formal) perturbative
expansion of $\Sigma$ in the Coulomb interaction $v$, yielding a series of
functionals of
$v$ and the non-interacting \GF\ $G_0$,
$\Sigma=vG_0+vG_0vG_0G_0+...$.
Alternatively, one can consider the self-consistent Dyson equation (SCDE) 
$G=G_0+G_0 S[G] G $, with $S$ the so called `self-energy functional',
giving rise to the self-consistent approach \cite{haussmann}.
Its formal justification can be given in terms of 
Legendre transformations \cite{dedominicis1964}, while the reference computable scheme
is the one proposed by Luttinger and Ward \cite{LW1960},
in which the self-energy functional is explicitly built as an infinite sum of
Feynman diagrams (sometimes referred to as the `skeleton series') obtained 
by formal manipulation of the perturbative expansion of the Green's function.

Due to the unknown properties of the perturbation expansion, the Luttinger-Ward approach
has not found solid mathematical foundations yet.
In fact, recent
numerical tests performed by Kozik and collaborators
on various Hubbard models \cite{kozik} showed clear signs of its failure in regimes 
of strong interactions.
Their results were soon connected to earlier numerical evidence of
possible pathologies in the diagrammatic approach by Sch\"afer and
 collaborators \cite{schafer2013,schafer2016}.
First investigations on an entirely analytical level were carried out by Stan and collaborators \cite{stan}, and Rossi and Werner \cite{rossi}, who independently managed to qualitatively reproduce the results of Kozik \etal 
using two analytically treatable toy
models. Those models, however, only bear the algebraic structure of the original quantum
mechanical problem and cannot be directly related to a Hamiltonian.
It would be then desirable to have a less system-specific view 
on the general problems of the self-consistent approach on one side, 
and a complete, analytical treatment of at least one of the particular Hamiltonians used 
in the work of Kozik \etal on the other side. 
This is the double goal of our work. 

In order to achieve this goal, we first recall and discuss the conditions 
under which the self-consistent approach can be set in general. 
This is done in Section \ref{sSCA}, where we briefly review
the current understanding of the foundations of the self-consistent approach
and its connection to its practical realization proposed by Luttinger and Ward. 
We then concentrate in Section II on the Hubbard atom, 
which is one of the systems studied by Kozik \etal \cite{kozik}.
In order to arrive at an analytical treatment, which makes a detailed and 
unambiguous analysis possible, we introduce functionals that realize 
the maps between the non-interacting and the interacting Green's functions 
\textit{on the physical domain of this specific model}. 
It is very important indeed to make a clear distinction
between the maps and the functionals that are used to realize the maps.
In particular, the numerical results of \cite{kozik} are understood 
as being due to a limitation of the skeleton series and 
the connected problem of the definition of the physical domain, 
whereas it remains still possible to build a valid self-energy 
functional and use the self-consistent approach.

More in detail,
after the discussion in Section I,
the presentation of the results and analysis for the Hubbard
atom in Section II is organized as follows: 
after an introduction (\ref{ssI}) we present the Hubbard atom (\ref{ssHA}),
prove the one-to-one correspondence between the $G_0$'s and $G$'s arising in this
model (\ref{ssMG0G}),
present explicit formulas for the functionals realizing those maps (\ref{ssEF}),
and use those to solve the SCDE (\ref{ssSCDE2}) and the inverse problem of finding
$G_0$ given $G$ (\ref{ssIP2});
then we define the `one-frequency--skeleton' series (\ref{ssLSKS}), which, in
analogy to the standard skeleton series,
when evaluated at the exact $G$, converges to the correct self-energy only in a 
subregion of the entire parameter space;
finally, we introduce the `one-frequency--SIN' series (\ref{ssLSS}),
which converges to the correct self-energy wherever the one-frequency--skeleton does
not,
 therefore offering the possibility to maintain the self-consistent approach 
for any interaction strength.
Conclusions will be drawn at the end.

%
%

\section{The Self-Consistent Approach}\label{sSCA}

\subsection{\GF s and self-energies}
\label{secWhy}

While DFT offers a computationally cheap way to get a good estimate of the charge
density
and several other observables, the lack of a known systematic way to connect the
density to any given 
observable represents a serious limit of the method for those who are interested in
other quantities.
On the other hand, the \GF\ formalism profits from Perturbation Theory, which 
gives a systematic, although expensive, way to write any N-body Green's function, 
and hence any observable, in terms of the non-interacting one-body \GF\ $G_0$ and the
interaction, at least at a formal level.
For instance, the one-body \GF\ (from now on, simply `\GF') is written,
in a simplified notation (see footnote 1), as
\be\label{Gper}
G=G_0+G_0vG_0G_0+G_0vG_0G_0vG_0G_0+...
\ee

The above expansion must be regarded
as a formal expression, and several problems may occur in practice. In particular,
the series may have a finite radius of convergence, 
        and hence not be suitable for strongly interacting regimes;
worse yet, it may have zero radius of convergence 
        and be asymptotic at best (see \cite{fischer1997} and references therein).
        Furthermore, the fact that we can write (\ref{Gper}) does not guarantee that also
    non-perturbative effects can be written in terms of $G_0$ only,
    for more information may be required.

It has been recognized early that at least some of the terms must be 
summed to infinite order to avoid divergent behavior. 
In particular, in the homogeneous electron gas 
one has to sum all so-called bubble diagrams, 
which express the polarizability of the system \cite{gellmann1957}.
Along this line, one widely used way to incorporate at least some of the diagrams
to all orders
is to recast (1) in the form of a Dyson equation \cite{fetter}
\be\label{De}
G=G_0+G_0\Sigma G
\ee
in which $\Sigma$, the `self-energy', is calculated via its perturbation expansion
\be\label{sigmaper}
\Sigma = vG_0+vG_0G_0vG_0+...
\ee

A further step on this route is the \emph{self-consistent approach},
based on the `self-consistent' Dyson equation (SCDE) in which the self-energy is
substituted
with a functional of the Green's function:
\be\label{scde}
G=G_0+G_0 S[G] G,
\ee
This approach has been put forth by Luttinger and Ward \cite{LW1960},
who proposed a formal expression for $S$ by manipulation
of the perturbative series (what we call the \emph{Luttinger-Ward approach}).
The resulting functional is itself in the form of a series, 
sometimes called `skeleton series', which we denote by
\be\label{Slw}
S_{LW}[G]\equiv vG+vGGvG+...
\ee
The skeleton series has the property of naturally leading to approximations to $S$ that
fulfill conservation laws for particle number, momentum and energy
\cite{Baym1961,PhysRev.127.1391},
which (\ref{sigmaper}) would not guarantee.
Moreover, since it involves the fully interacting $G$, rather than $G_0$,
one may hope that (\ref{Slw}) has a faster convergence than (\ref{sigmaper}),
leading in turn to better approximations to $G$ than (\ref{sigmaper})
when truncated at the same order.
However, apart from a few further developments \cite{dedominicis1964},
very little is known about the consistency of the Luttinger-Ward approach. 
In fact, the recent results of Kozik and collaborators \cite{kozik}
provide numerical evidence for its failure for certain regimes.

\subsection{On the Self-Consistent Dyson Equation}\label{ssITSCARC}

Being derived within Perturbation Theory, the Luttinger-Ward approach
formally applies to all quantum field theories.
However, neither the original perturbative derivation, nor further non-perturbative
developments fully guarantee that the self-consistent approach
always leads to the \GF\ of the original Hamiltonian problem. 
For this to happen, four conditions have to be fulfilled:
\begin{enumerate}
 \item given a certain interaction, all information about the specific system of
interest 
           is encoded in $G_0$, the sole input of (\ref{scde});
\item given $G$ and $G_0$, there exists a
$\Sigma$ such that 
          $G=G_0+G_0\Sigma G$ holds;
\item there exists a functional $S$ such that $S[G]=\Sigma$.
\end{enumerate}
Those three conditions guarantee that $X=G$ is solution of the equation
\be\label{scdex}
X=G_0+G_0 S[X]X.
\ee
However, the equation might have other, spurious solutions,
which could be either an excited state solution, a Green's function of another system, 
or a function that does not correspond to any `physical' Green's function.
This would not threaten the theoretical foundations of the approach,
it would, however, represent a serious obstacle for its practical use.
Therefore we formulate a forth condition:
\begin{enumerate}[resume]
 \item the functional $S$ should be such that (\ref{scdex}) has only one solution,
namely $X=G$.
\end{enumerate}
In the following, we shall make these four conditions more precise
and put them in the context of current knowledge.

First, we need to give a definition of \emph{physical} Green's function.
Given an interacting Hamiltonian, the $N$-body Green's functions are
unambiguously defined as $2N$-point correlators. We focus on the zero
temperature, equilibrium one-body Green's function,
\be\label{Gcorr}
G(1,2)\equiv -i\langle \Psi|\mathcal{T}[\hat{\psi}^\dagger(1)\hat{\psi}(2)]|\Psi\rangle,
\ee
where $|\Psi\rangle$ represents the ground-state of the Hamiltonian at
fixed particle number, $\mathcal{T}$ the time-ordering operator, 
$\hat{\psi}$ the field operator in the Heisenberg representation,
and $1,2$ compact arguments representing space, spin and time degrees of freedom, 
$1\equiv({\bf r}_1,\sigma_1,t_1)$.
The definition (\ref{Gcorr}) does not depend on the chosen Hamiltonian.
A specific $G$ is obtained when the Hamiltonian, in its parametric form, is decided, 
specific values for those parameters are chosen, and all information
required to fully identify the state (in particular the number of particles) is given.
When dealing with the electronic many-body Hamiltonian, we say that all `systems' are
characterized by the same interaction (the Coulomb interaction) and 
they are distinguished by the external potential.
We generalize this notion in the following way.
Given any (parametric) Hamiltonian, different \emph{systems} are obtained by varying
the value of the parameters appearing in the non-interacting part,
while keeping the value of the interaction fixed.
This allows us to introduce the notion of `physical Green's function' as follows.
Given a certain Hamiltonian with \emph{fixed} value of the interaction,
the set $\{G\}$ of physical Green's functions is represented by all and only the
functions 
obtained by evaluating formula (\ref{Gcorr}) on all possible \emph{systems}.\footnote{%
	Since the set may change with the value of the interaction,
	a more precise notation would be $\{G\}_v$.
	Working always at fixed $v$, however, we shall omit the subscript $v$,
	for sake of notational economy.}
In the out-of-equilibrium case an analogous definition holds,
while, in case of finite-temperature, one should take into account that
different systems are also characterized by different values of temperature.
Obviously, there is no guarantee that any arbitrary function of two (multi-dimensional) arguments $f(1,2)$ belongs to
$\{G\}$,
i.e. that it corresponds to the Green's function of some system.
Moreover, the set $\{G\}$ may change according to the interaction and the type of
Green's function 
(zero/finite temperature, in/out of equilibrium) considered.
Finally, we note that here we consider the case of the electronic many-body problem, 
in which the interaction is fixed once for all to Coulomb,
but our discussion would not change if a different choice were made.

In the following we shall refer to elements of $\{G\}$ as `physical Green's functions'
or, simply, `Green's functions', all other functions of two arguments being `unphysical
Green's functions'.
In an equivalent fashion, the set of `physical \textit{non-interacting}
Green's functions' $\{G_0\}$ can be established.
It should be emphasized that this is an important difference with other works,
like \cite{kozik}, in which this distinction is not made, and the term `Green's
function' seems to 
include any function of two arguments, covering both `physical' and `unphysical' ones.

This definition allows us to reformulate the first point of our list of conditions in
the following way:
given a certain Hamiltonian (with fixed value of the interaction,
but all possible values of its parameters), the corresponding set
$\{G_0\}$ must identify
all elements of the set $\{G\}$ completely and without ambiguities.
In other words, the map $\{G_0\}\to\{G\}$ must be surjective.
\begin{figure}[t]
\includegraphics[width=5.5cm]{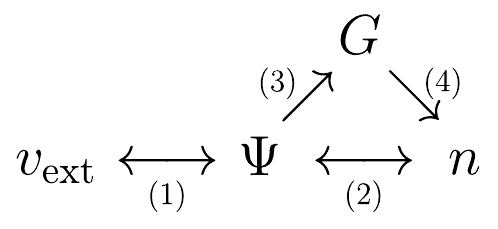}
\caption{As mentioned in \cite{stan}, for zero temperature, 
equilibrium Green's functions $G$,
the Hohenberg-Kohn theorem, which connects the external potential $v_{\rm ext}$
to the corresponding (non-degenerate) ground-state wavefunction $\Psi$ (link 1) and to the density
$n$ (link 2),
guarantees that there is a one-to-one 
correspondence between $v_{\rm ext}$ and $G$, in the following way:
if we know $v_{\rm ext}$, we can build $\Psi$ (link 1) and hence $G$ (link 3);
on the other hand, if we know $G$ we then know $n$ (link 4) and hence $v_{\rm ext}$
(links 2,1). Since such a correspondence $\{v_{\rm ext}\}\leftrightarrow\{G\}$
holds also in the non-interacting case $\{v_{\rm ext}\}\leftrightarrow\{G_0\}$,
it follows that $\{G_0\}\leftrightarrow\{G\}$.
}\label{figDFT}
\end{figure}



The second point concerns the existence of the self-energy $\Sigma$.
Formally, one can write $\Sigma=G_0^{-1}-G^{-1}$,
where the inverse functions $G^{-1}$ and $G_0^{-1}$,
from now on collectively shorthanded with $G_{(0)}^{-1}$, are defined by
\begin{multline}\label{Ginv}
\int d3 G_{(0)}(1,3)G_{(0)}(3,2)^{-1}=\\
=\int d3 G_{(0)}(1,3)^{-1}G_{(0)}(3,2)=\delta(1,2).
\end{multline}
For a self-energy to exist it is therefore enough that $G_{(0)}^{-1}$ exists.

The third issue is the existence of a functional $S$ with $S[G]=\Sigma$.
For this we need that $\{G_0\}\to \{G\}$ is also injective,
which makes $\{G_0\}\leftrightarrow\{G\}$ a one-to-one map. 
This indeed would allow us to build a functional $L$ such that $L[G]=G_0$
and $S$ as $S[f]\equiv L[f]^{-1}-f^{-1}$, 
where the inverse $L[f]^{-1}$ is guaranteed to exist for at least $f=G$ by the
existence
of the inverse of $G_0$.\footnote{%
	Here inverse $L[f]^{-1}$ is again understood as in (\ref{Ginv})
	and not as the `inverse functional', denoted by $L^{-1}[f]$, for which
	$L^{-1}[L[f]]=f$.}

Concerning the non-relativistic electronic problem in particular,
because of the the simple relation connecting the charge density to the
Green's function $n(1)=-iG(1;1^+)$,
one can use theorems developed in the framework of Density
Functional Theory \cite{hohekohn,kohnsham,Mermin1965,rungegross} to establish the
relation
between $\{G_0\}$ and $\{G\}$, which, in DFT wording, are the
set of all $v-$representable (non-interacting and interacting, respectively) Green's functions.
As one can see in fig. \ref{figDFT}), a simple argument 
points indeed to a one-to-one map between the two sets, at
least in the zero-temperature equilibrium case.

\begin{figure}[t]
\includegraphics[width=7.5cm]{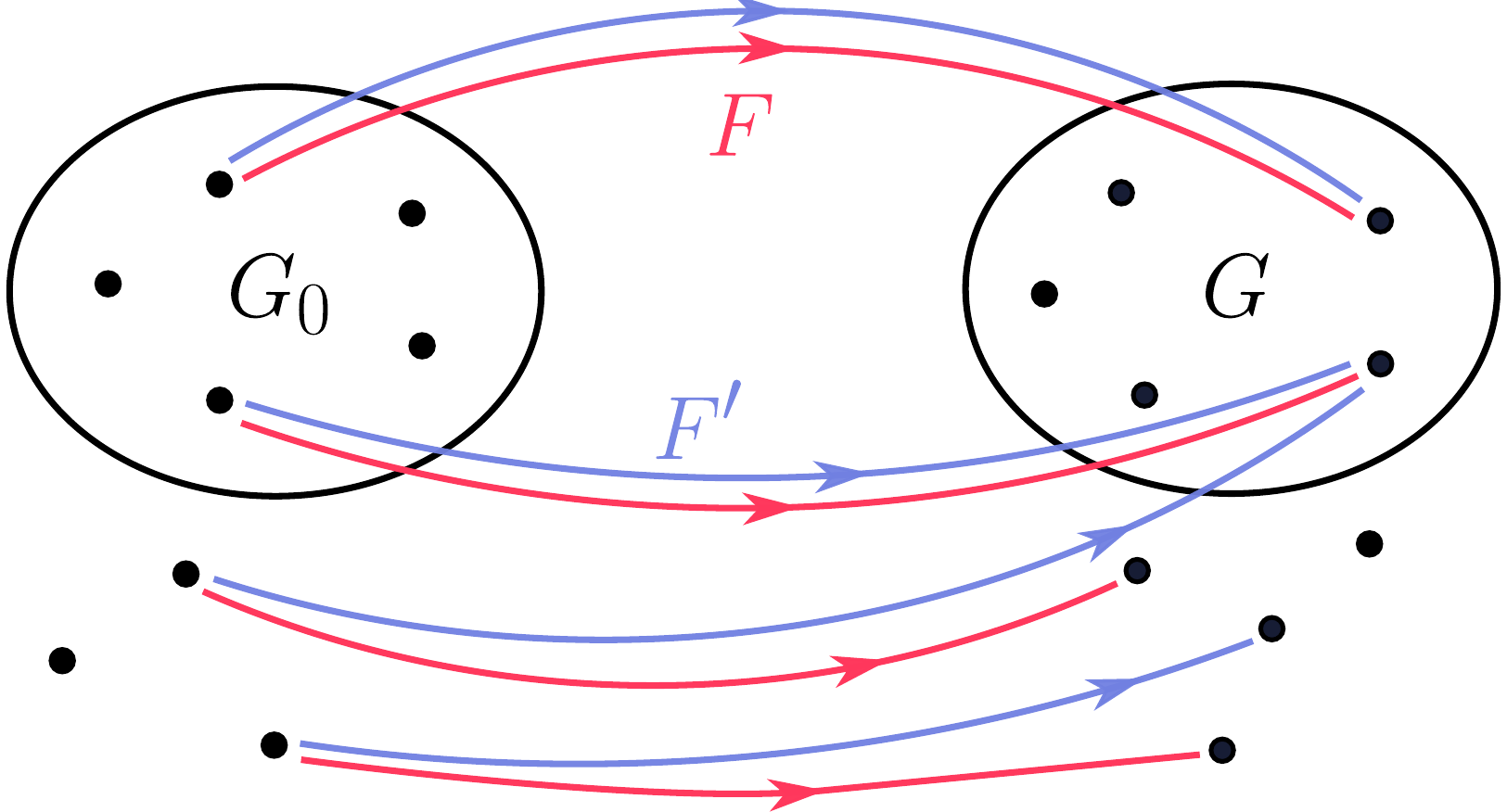}
\caption{%
Pictorial representation of two functionals $F$ and $F'$
that correctly realize the map $\{G_0\}\to\{G\}$, but have
different behavior outside the set $\{G_0\}$.
In particular, in the present example $F$ is injective ($F[f]\neq
F[g]$ for all $f,g$ in its domain)
while $F'$ is non-injective (there is at least one $G_0$ and one
$f\notin \{G_0\}$
for which $F'[G_0]=F'[f]$).}\label{map1}
\end{figure}

\begin{figure}[t]
\includegraphics[width=7.5cm]{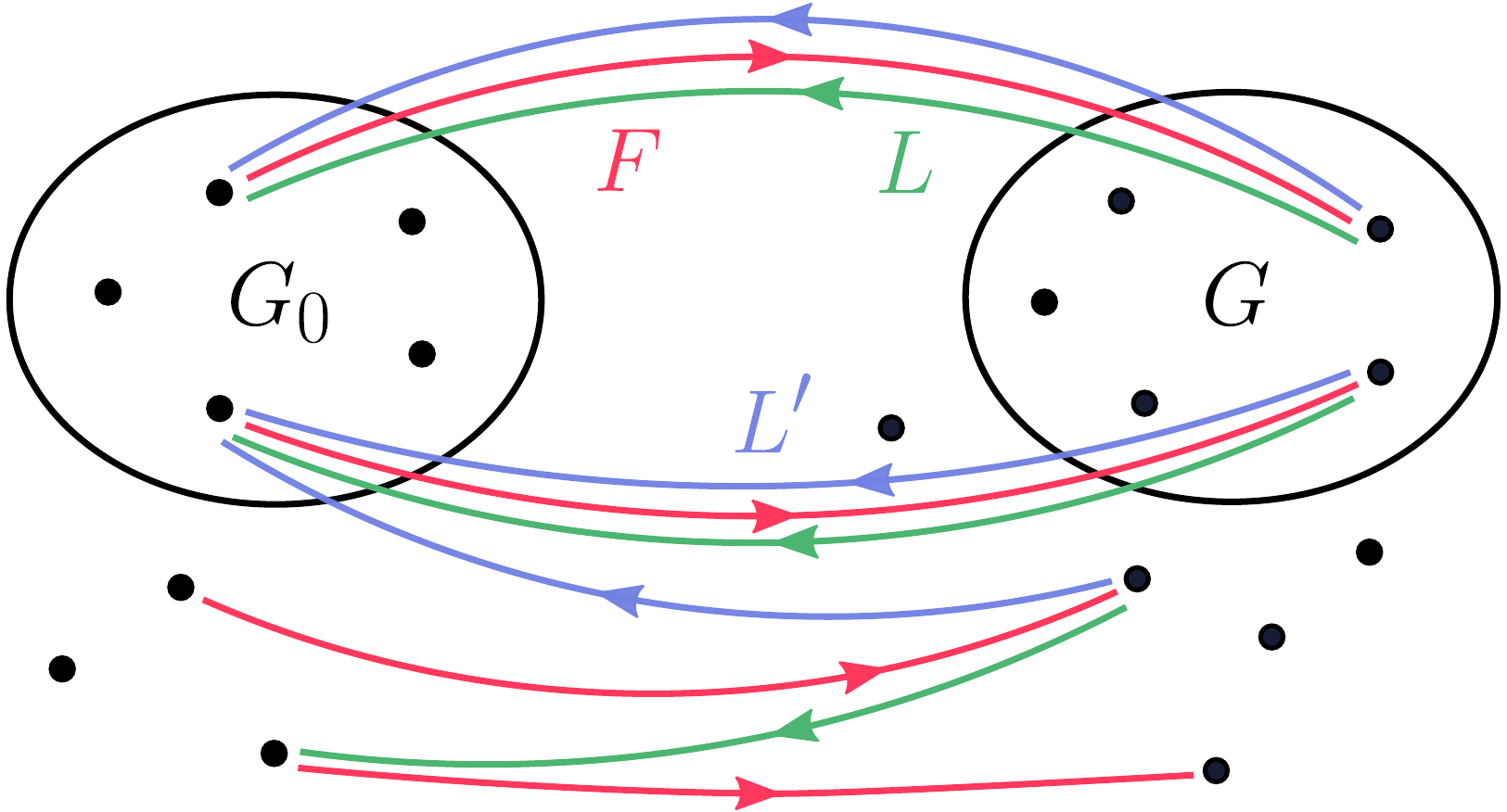}
\caption{%
While $F$ realizes the map $\{G_0\}\to\{G\}$, $L$ and $L'$
realize the opposite map $\{G_0\}\leftarrow\{G\}$. 
$L$ is injective on its entire domain, while $L'$ is not.
Both $L$ and $L'$ are equal to $F^{-1}$ on the domain limited to
$\{G\}$,
but they are different from $F^{-1}$ when the entire domain is
allowed.}\label{map2}
\end{figure}

\begin{figure}[t]
\includegraphics[width=7.5cm]{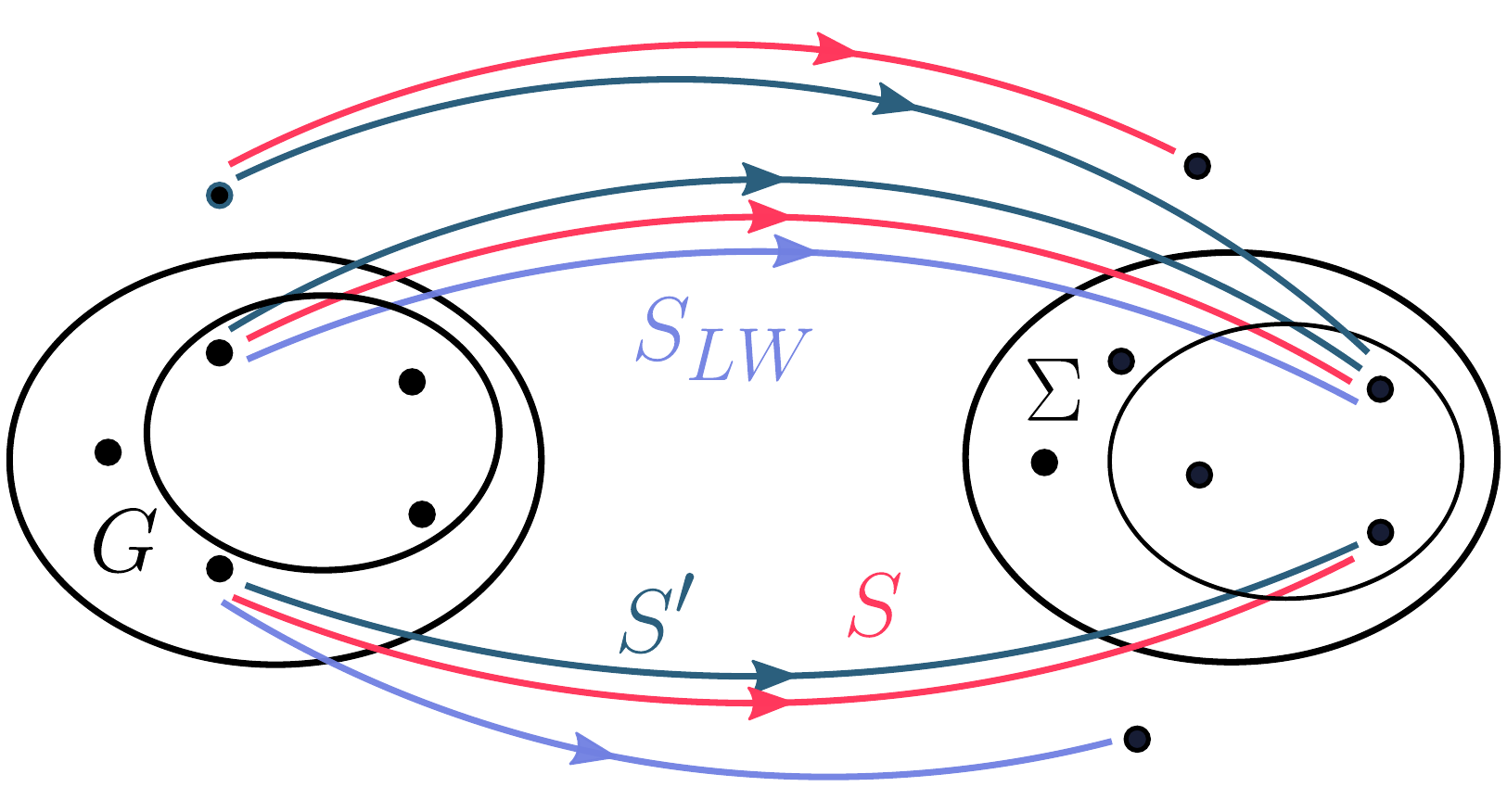}
\caption{%
Left outer set: physical Green's functions; 
right outer set: physical self-energies.
$S$ and $S'$ are two distinct functionals for which $S[G]=S'[G]=\Sigma$.
$S_{LW}$ represents the skeleton series introduced by Luttinger and Ward.
It satisfies $S_{LW}[G]=\Sigma$ but not for all physical $G$'s.
The equality holds only for the left inner set mapped to the right inner set.}\label{map3}
\end{figure}

Finally, given a functional $S$ such that $S[G]=G_0^{-1}-G^{-1}$,
we have to distinguish two situations. If the domain of $S$ is $\{G\}$,
then the above conditions ensure that the SCDE has one and only one solution.
If the domain of $S$ extends beyond $\{G\}$ then a problem of multiple,
spurious solutions might occur.
The skeleton series, for instance, takes as entry any function of two arguments,
unless specified differently.
Since in the case of the electronic many-body problem $\{G\}$ does not cover
the entire space of functions with two arguments,\footnote{%
	For instance, $-iG(1,1^+)$ corresponds to the charge density, which cannot be negative.}
the SCDE with the skeleton series could have spurious solutions, 
even when the series, evaluated at a certain $G$, gives the correct $\Sigma$.
To better understand this situation, let us discuss a few subtleties.
First, consider the map $\{G_0\}\to\{G\}$, for which there exists a functional,
say $F$, that takes elements of $\{G_0\}$ to the corresponding element of $\{G\}$, 
or, in formulae, $F[G_0]=G$.
If $\{G_0\}$ is only a proper subset of all possible functions of two arguments,
then $F$ is not the only functional that realizes that map, 
for another functional $F'$ such that $F'[G_0]=G$ and yet $F[f]\neq F'[f]$
if $f\notin\{G_0\}$ can exist (fig. \ref{map1}).
Since the map is supposed to be one-to-one there must also be a functional or,
better, a
family of functionals realizing the inverse map, namely $L[G]=G_0$.
Given two functionals such that $F[G_0]=G$ and $L[G]=G_0$, we cannot conclude
that $L=F^{-1}$, since it might be that $L[f]\neq F^{-1}[f]$ if $f\notin\{G\}$
(see fig. \ref{map2}). 
Different functionals $L$ implementing the map $\{G\}\to\{G_0\}$ would obviously
lead to different $S$
functionals, as shown in fig. \ref{map3}.
For $S[f]=L[f]^{-1}-f^{-1}$ 
the equation $X=G_0+G_0S[X]X$ is equivalent to
 $L[X]=G_0$.
Therefore, it has  
only one solution in the set $\{G\}$
but can have other solutions outside that set, the number and nature depending
on the features of the chosen $L$.
Those spurious solutions can be avoided in two ways:
either we build $S$ on an $L$ that is injective on its entire domain
(i.e. there are no two $f$ and $g$ for which $L[f]=L[g]$)
or we restrict the search for solutions to the set $\{G\}$.
Unfortunately, neither option is easy to realize, the first one for a lack
of methods to build $L$ with the desired properties and the second for a lack
of a complete characterization of the set $\{G\}$.\footnote{%
	For instance, although the analytic structure of the equilibrium Green's function
	in frequency space is known to be a sum of poles \cite{lehmann1954}
	exact constraints on coefficients of these poles are only partially known.
	In fact, even for the charge density, which is a part of the \GF,
    the problem of finding all necessary constraints 
     (in this context called $v-$representability)
    remains unsolved (see \cite{engeldreizler2011} and references therein).}

\subsection{On the Inverse Problem}\label{ssIP}

If $\{G_0\}\leftrightarrow\{G\}$ holds and we are given a functional $F$
such that $F[G_0]=G$, one may want to use it for solving the inverse
(or `embedding') problem: $F[X]=G$, in which $G$ is given and $G_0$ is to be found.
In general, $F$ can be non-injective over its entire domain, 
and one can run into spurious solutions (see $F'$  in fig. \ref{map1}).
Unless the domain of $F$ is properly restricted,
this may happen, for instance, if one attempts to solve the Dyson equation
with a self-energy evaluated in perturbation theory (eqn (\ref{De}-\ref{sigmaper})).
On the other hand, if the functional $F$ relies on solving a SCDE,
then the corresponding inverse problem has always only one solution, the physical $X=G_0$, 
irrespective of the properties of the $S$ appearing in the SCDE and of possible
restrictions
on the search for solutions.
This is because the SCDE as an equation in $G_0$, rather than $G$,
namely $G=X+XS[G]G$, is a \emph{linear} equation whose unique solution
is $X=(S[G]+G^{-1})^{-1}$ which readily simplifies to
$X=(L[G]^{-1}-G^{-1}+G^{-1})^{-1}=L[G]=G_0.$

\section{Self-Consistent Approach to the Hubbard Atom}\label{sSCAHA}
\subsection{Introduction}\label{ssI}

The previous section has set the general frame. In particular, it emerges that  
in order to avoid spurious, unphysical solutions to the SCDE, 
one must either use a self-energy functional $S[f]\equiv L[f]^{-1}-f^{-1}$ with $L$
injective over its entire domain,
or restrict the search of solutions to the set $\{G\}$ where $L$ \textit{is}
injective; physical \GF s must lie within this set.

In order to illustrate these quite abstract considerations, and in order to use them
to explain current problems in the 
literature and search for potential solutions,
we are going to analyze the self-consistent approach
on a simple, solvable model: the finite temperature, half-filled Hubbard Atom which
is defined in the next subsection.

The Hubbard atom has been one of the systems studied in the 
recent work by Kozik and collaborators \cite{kozik}. In their Letter 
the outcome of several numerical tests performed on Hubbard models was reported.
One set of tests is devoted to probe the number and nature of solutions of
the inverse problem, defined in Section \ref{ssIP}.
They found one physical and one unphysical solution. 
The remaining tests concern the skeleton series, which, when evaluated at the exact $G$,
converges to the correct self-energy only for weakly interacting systems.
On the basis of these numerical results, two main conclusions were proposed: 
{\rm (i)} the map $G_0 \to G$ is not invertible;
{\rm (ii)} the skeleton series has two branches, and one may talk about the 
``non-existence of the Luttinger Ward functional''.

In the following
(subsection \ref{subsec:map}) we shall prove that, at least in the Hubbard atom,
the map $\{G_0\}\leftrightarrow\{G\}$ holds. 
This will establish the self-consistent approach as an exact and closed rewriting 
of the original problem to determine the \GF\ given the Hamiltonian.
Furthermore, we shall present an actual realization of the approach.
We will first present two explicit functionals $F_W$ and $L_W$ that, limited to this
model,
realize the map $\{G_0\}\rightarrow\{G\}$ and $\{G_0\}\leftarrow\{G\}$, respectively.
The functional $L_W$ will then be used to build a functional $S_W$ that allows to
write down
the SCDE in an explicit form. We shall then prove that such a SCDE has one and only
one solution in the set of physical Green's functions.
On the other hand, the functional $F_W$ will be used to show that the inverse problem
$F[X]=G$ has spurious solutions if $F$ is non-injective and the search of solutions
is not restricted to the set of physical non-interacting Green's functions.
Moreover, the formal perturbative expansion of $F_W$ will allow us to build
the `one-frequency--skeleton' series, in analogy to the standard skeleton series.
The one-frequency--skeleton series evaluated at $G$ will be shown to converge to the
correct
self-energy only in a subregion of the parameter space.
Finally, we shall construct a second series (`one-frequency--SIN')
that \emph{complements} the one-frequency--skeleton, in the sense that, when
evaluated at $G$,
it converges to the correct self-energy wherever in the parameter space 
the one-frequency--skeleton does not, and vice versa.


\subsection{The Hubbard Atom}\label{ssHA}

The Hamiltonian of the Hubbard atom is
\be\label{hamiltonian}
\hat{H}= \frac{U}{2}\sum_{\sigma,\sigma'=\uparrow,\downarrow}\hat{c}^\dagger_\sigma
\hat{c}^\dagger_{\sigma'} \hat{c}_{\sigma'} \hat{c}_\sigma,
\ee
where $\hat{c}_\sigma^{(\dagger)}$ is a fermionic annihilation (creation) operator
and $U>0$ is the interaction parameter.
In order to establish a close connection to \cite{kozik}, we calculate 
the finite temperature one-body Green's function in the grand canonical ensemble,
which is defined as \cite{fetter71}
\be\label{Gdefinition}
G_{\alpha\beta}(\vec{x},\tau;\vec{x}',\tau') =
	\frac{\mathrm{Tr}\left\{e^{-\beta(\hat{H}-\mu\hat{N})} \mathcal{T}[\hat{\psi}_\alpha^\dagger(\vec{x},\tau)\hat{\psi}_\beta(\vec{x}',\tau')]
\right\}}{\mathrm{Tr}\left\{e^{-\beta(\hat{H}-\mu\hat{N})} \right\}}
\ee
where $\hat{N}$ is the number operator, $\mu$ is the chemical potential, 
$\mathcal{T}$ orders operators according to their value of $\tau$,
$\hat{\psi}_\alpha(\vec{x},\tau)$ is the field operator and
$\beta=1/(k_B T)$ with $T$ the temperature and $k_B$ the Boltzmann constant. 
The frequency Fourier transform to imaginary
frequencies yields the Matsubara \GF\  \cite{fetter71}.
For the Hubbard atom (\ref{hamiltonian}) and for a chemical potential $\mu=U/2$ this
yields $\mathcal{G}_{\sigma \sigma'}(z)=\G\delta_{\sigma\sigma'}$ with
\begin{eqnarray}
 \label{Gdef}
\G&\equiv& \frac{1}{2}\left(\frac{1}{z+U/2}+\frac{1}{z-U/2}\right)\\
&=&\frac{1}{z}\frac{1}{1-U^2/(4z^2)}
\end{eqnarray}
with 
\begin{equation}
 z\equiv i\omega_n\,\,\,\,\,\,\,\,\,\,\,\textrm{and}\,\,\,\,\,\,\,\,\,\,\,\,\omega_n  \equiv k_B\pi T(2n+1), \,\,\,\,\,\,n\in \mathbb{Z}.
 \label{eq:definition}
\end{equation}
The corresponding non-interacting Green's function is obtained by setting $U$ to 0,
which yields 
\begin{equation}\label{gdef}
 \g\equiv \frac{1}{z}.
\end{equation}
The self-energy is then defined as 
\be\label{SigmaAtom}
\Sigma(z)\equiv \g^{-1}-\G^{-1}=\frac{U^2}{4z}.
\ee

\subsection{The map  \texorpdfstring{$\{G_0\}\leftrightarrow\{G\}$}{Gg}}\label{ssMG0G}
\label{subsec:map}

In a Hubbard model with $N>1$ sites at finite temperature, what identifies a
specific system,
and hence the corresponding Green's function, is the value of the temperature $T$,
the chemical potential $\mu$ and the value of the hopping expressed in units 
of the interaction strength $U$ as $t/U$. Sometimes the hopping parameter $t$ is fixed to 1
and $U$ let to vary. However, in case of one site, no hopping is possible, so $U$ is simply
a \emph{constant} that sets the energy scale.
Moreover, in our case the chemical
potential
is also fixed, leaving the temperature $T$ as the only parameter that really
characterizes
a system.
We then say that the set of physical Green's functions $\{\G\}$ is defined as the set
of all and only the functions in $n\in \mathbb{Z}$ obtained from (\ref{Gdef}) and (\ref{eq:definition}) by
varying
$T\in(0,\infty)$. In other words, any function of $n$ that cannot be written as
(\ref{Gdef})
for a certain value of $T$ will be considered an `unphysical' Green's function.
The set of physical non-interacting Green's functions $\{\g\}$ can then be defined in
an analogous way.

The parametrization of the two spaces in terms of the temperature $T$
is quite convenient in view of the analysis of the 
map between them.
In fact, if we define the space of physical temperatures 
$\{T;T> 0\}$, the one-to-one--ness of the map
$\{\g\}\leftrightarrow \{\G\}$
can be proved as a consequence of the one-to-one--ness of the maps
$\{T\}\leftrightarrow \{\g\}$
and $\{T\}\leftrightarrow \{\G\}$, as follows.

Let us first look at $\{T\}\leftrightarrow \{\g\}$.
Since we have that physical $G_0$'s are defined starting 
from physical temperatures (i.e. for any physical temperature
there is one physical $\g$: $\{T\}\rightarrow \{\g\}$), 
we need to prove that every physical $T$ corresponds to only one $\g$.
If $T'$ and $T$ gave rise to the same $\g$ we could write
\be\label{gT}
\frac{1}{i(2n+1)\pi k_B T}=\frac{1}{i(2n+1)\pi k_B T'}.
\ee
But this simply reduces to $T=T'$, so there are no two physical temperatures
leading to the same physical non-interacting Green's function.

Concerning the map $\{T\}\leftrightarrow \{\G\}$,
we have that $\{T\}\rightarrow \{\G\}$ is again realized by definition,
while for the inverse map we must solve the equivalent of (\ref{gT}) 
which reads:
\begin{multline}\label{oneeq}
\frac{1}{2}
        \left(\frac{1}{i(2n+1)\pi k_B T'+\frac{U}{2}}
                +\frac{1}{i(2n+1)\pi k_B T'-\frac{U}{2}}\right)
=\\
=\frac{1}{2}
        \left(\frac{1}{i(2n+1)\pi k_B T+\frac{U}{2}}
                +\frac{1}{i(2n+1)\pi k_B T-\frac{U}{2}}\right)
\end{multline}
which \textit{per se} has two solutions:
\be\label{twosol}
T'=T\;\;\;\text{and}\;\;\;T'=\frac{U^2}{4 \pi ^2 k_B^2 (2 n+1)^2 T}
\ee
However, the second one is not a valid physical temperature,
since it is a function in $n$ rather than a positive number.
This means that a physical interacting Green's function
corresponds to only one physical temperature.

Since $\{T\}\leftrightarrow \{\g\}$ and
$\{T\}\leftrightarrow \{\G\}$ holds, one can then conclude that 
$\{\g\}\leftrightarrow \{\G\}$.

\subsection{Explicit Functionals}\label{ssEF}

Both sets $\{\g\}$ and $\{\G\}$ only represent portions of 
all possible functions in $z$.\footnote{%
	We intentionally leave 
	open the definition of what \emph{`all possible functions'} is.
	All following statements are valid, for instance, if one considers 
	the set of all rational functions or the bigger set of analytic functions.
	What matters is only to define the domain of the functionals one considers to be the physical domain.}
It follows that there can be different functionals realizing the maps
$\{\g\}\leftrightarrow\{\G\}$,
characterized by inequivalent results outside those sets.
A specific functional realizing the map $\{\g\}\to\{\G\}$ is\footnote{%
    $F_W$ has the same structure as the ones studied by
    Rossi \etal and Sch\"afer \etal \cite{rossi,schafer2016}
    and first introduced by Stan and collaborators
    \cite{stan} in the context of the so called One Point Model (OPM)
    \cite{Lani2012,berger2014},
    which is a mathematically simplified framework for studying the 
    functional formulation of the Green's function formalism.
    (see also \cite{molinari2005,pavlyukh2007}).}
\be\label{FW}
F_W(n,[f])\equiv\frac{f(n)}{1-f(n)^2\frac{U^2}{4}}.
\ee
This functional takes as input any function $f(n)$, not just non-interacting
Green's functions.
A functional realizing the inverse map $\{\g\}\leftarrow\{\G\}$ is

\be
L_W(n,[f])\equiv\left(i(2n+1)\frac{U}{2}\sqrt{-\frac{\frac{1}{3}
        +\frac{f(-1)}{f(1)}}{3+\frac{f(-1)}{f(1)}}}\right)^{-1}.
        \label{eq:def-lw}
\ee
Note that $L_W$ carries an explicit dependence on $n$, while $F_W$ does not.
For economy of notation, in the following we shall denote all functionals by $\mathcal{O}[f]$,
rather than $\mathcal{O}(n,[f])$, omitting to indicate a possible dependency on $n$.

The functionals $F_W$ and $L_W$ are different from the corresponding ones built in Perturbation Theory.
They have restricted capabilities, for they give the correct result only when applied to Green's functions
of the Hubbard atom, viz. (\ref{Gdef}) and (\ref{gdef}). 
A difference that will play a central role in our later discussion 
(Sections \ref{ssIP2} to \ref{ssLSS}) is that $F_W$ is completely \emph{local} 
in (Matsubara) frequency space 
(the value of $F_W[G]$ at a specific $n=n_0$ depends only on $G(n')$ at that $n'=n_0$),
while Perturbation Theory prescribes convolutions on frequencies.
On the other hand, $L_W$ is characterized by a universal (i.e. $f$-independent) function,
namely $(2 n+1)$, which encodes the structure common to all $\g$'s, 
and a prefactor that depends on $f$ in a non-local way,
which, evaluated at $f=\mathcal{G}$, 
extracts the information about the specific system.

\subsection{Self-Consistent Dyson Equation}\label{ssSCDE2}

\begin{figure}[t]
\includegraphics[width=8.5cm]{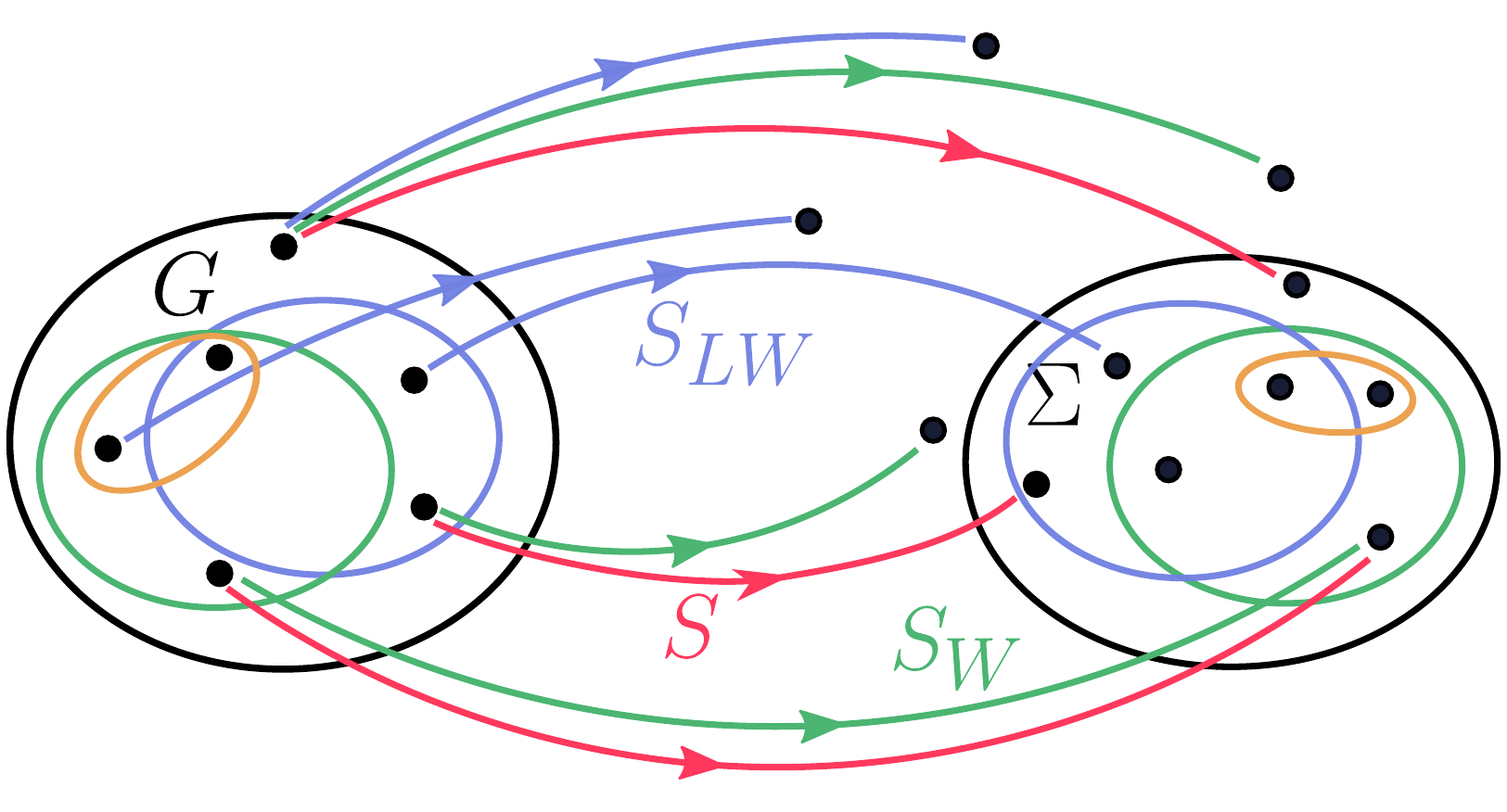}
\caption{%
A self-energy functional $S$ realizes the map $\{G\}\to\{\Sigma\}$
for all physical $G$'s (black sets).
Just like the skeleton series $S_{LW}$ satisfies $S_{LW}[G]=\Sigma$
only for certain $G$'s (blue subsets), so does $S_{W}$ for which $S_{W}[G]=\Sigma$
(green subsets). Contrary to $S_{LW}$,
the subset for which $S_W[G]=\Sigma$ (orange subsets) 
includes all $G$'s of the Hubbard Atom.}\label{figmap4}
\end{figure}

Using $L_W$ we can define the functional
\begin{multline}
S_W[f]\equiv L_W[f]^{-1}-f(n)^{-1}=\\
	=i(2n+1)\frac{U}{2}\sqrt{-\frac{\frac{1}{3}
        +\frac{f(-1)}{f(1)}}{3+\frac{f(-1)}{f(1)}}}-\frac{1}{f(n)}.
\label{eq:se}
\end{multline}
From that we can build the SCDE
\be\label{SCDESW}
X=\g + \g S_W[X] X
\ee
with $X$ an unknown function of $n$ (see fig. \ref{figmap4}).
As pointed out in Section \ref{ssITSCARC}, (\ref{eq:se}) with (\ref{SCDESW}) are equivalent to $\g = L_W[X] $.
Since $L_W$ is not injective
\footnote{%
	Consider for instance $L_W[X]$
	and $L_W[X + f]$ with $f$ any function for which $f(-1)=0$.},
given $\g$, this equation has multiple solutions. However, 
equation (\ref{SCDESW}) 
has only one solution if we restrict our search to the set $\{\G\}$.
To see this explicitly, we use the expression for $\g$:
\be
\frac{1}{i(2n+1)k_B\pi T}=L_W[X].
\ee
We enforce the correct domain on $X$ by considering only functions
belonging to $\{\G\}$. We therefore look at
\be\label{eqnLoff}
\frac{1}{i(2n+1)k_B\pi T}=L_W[f_x(n)]
\ee
with
\begin{multline}
f_x(n)\equiv \frac{1}{2}\left(\frac{1}{i(2n+1)\pi k_B x+\frac{U}{2}}+\right.\\
\left.+\frac{1}{i(2n+1)\pi k_B x-\frac{U}{2}}\right)
\end{multline}
where $x$ is an unknown temperature.
The evaluation of $L_W$ defined in (\ref{eq:def-lw}) on $f_x(n)$ leads to
\be
L_W[f_x(n)]=\frac{1}{i(2n+1)k_B\pi x}.
\ee
We can then write (\ref{eqnLoff}) as
\be
\frac{1}{i(2n+1)k_B\pi T}=\frac{1}{i(2n+1)k_B\pi x}
\ee
from which we can determine $x$ without ambiguities to be
\be
x=T.
\ee
Plugging this value back in $f_x(n)$ gives the final, correct answer:
\be
X=\frac{1}{2}\left(\frac{1}{i(2n+1)\pi k_B T+\frac{U}{2}}
        +\frac{1}{i(2n+1)\pi k_B T-\frac{U}{2}}\right).\nonumber
\ee

\subsection{Inverse Problem}\label{ssIP2}

\begin{figure}[t]
\includegraphics[width=8.5cm]{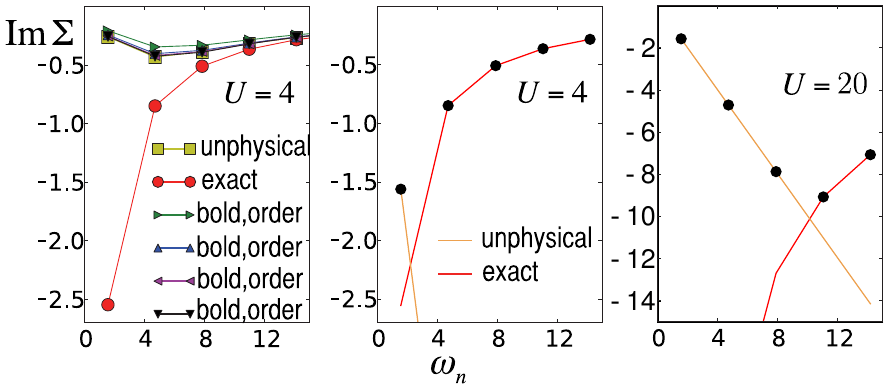}
\caption{%
Left panel: third panel of fig. 3 from \cite{kozik},
representing the physical (red) and the 
unphysical (black) self-energies corresponding to the physical and the unphysical
solutions to the inverse problem $F_{QMC}[X]=G$,
with $F_{QMC}$ the functional defined by the
Quantum Monte Carlo code used in \cite{kozik} to calculate $G$ from a given $G_0$.
Second and third panel: physical (red) and unphysical (orange) self-energies
corresponding to the physical and unphysical solutions to $F_W[X]=G$,
for two different values of $U$.}\label{figspurious}
\end{figure}

We now look at the problem $F_W[X]=\G$, in which $\G$ is known
and $\g$ is to be found.
The inverse problem has two solutions: the correct $X=1/z$
and a spurious one $X=-\frac{4z}{U^2}$, which we will denote by $\gu$.
Since there is no $T$ such that $\g$ can be written as $\gu$,
$\gu \notin \{\g\}$, $\gu$ is an unphysical solution, as expected.
The correct solution is therefore identified without ambiguity 
by restricting the domain to $\{\g\}$. If, like in \cite{kozik}, 
the domain is not properly restricted, 
the spurious solution survives, leading to a problem 
of `multiple solutions' \cite{stan,eder2015}.
As anticipated, number and nature of the spurious, unphysical solutions depend
on the specific functional one uses. 
Since our local functional is different from the functional of Kozik and collaborators,
we expect that our $\gu$ does not correspond to the unphysical solution 
found by these authors. Since \cite{kozik} provides no direct information on their spurious solution,
but only a plot with the imaginary part of the corresponding self-energy,
in fig. \ref{figspurious} we also plot the imaginary part 
of the self-energy arising from our spurious solutions,
namely $\Sigma_u(z)\equiv \gu^{-1}-\G^{-1}$. 
The comparison between the two self-energies shows
that indeed the two spurious solutions are different.


\subsection{Local-Skeleton Series}\label{ssLSKS}

We will now turn to the convergence of the skeleton series,
motivated by the numerical evidence provided by Kozik \etal
for the fact that the skeleton series may converge to a wrong functional.
An analytic proof and analysis for the full LW skeleton series is currently out of reach.
However, we can obtain and analyze a qualitatively equivalent result
by studying the behavior of the `one-frequency--skeleton' series, built as follows.

In standard Perturbation Theory, the skeleton series arises as a formal manipulation
of the terms of the expansion of the fully interacting Green's function in
terms of the non-interacting one \cite{LW1960}. One way to proceed
can be summarized as follows:
we first start with the perturbative expansion of $G$:
\be\label{Gper2}
G=G_0+v G_0^3+v^2 G_0^5+...,
\ee
We then use (\ref{Gper2}) to write $G_0$ in terms of $G$:
\be\label{gper}\begin{array}{l}
G_0=G-\left(v G_0^3+v^2 G_0^5+...\right)\\
\Rightarrow G_0=G-\left(v \left(G-\left(v G_0^3+v^2 G_0^5+...\right)\right)^3+v^2
G_0^5+...\right)\\
\Rightarrow G_0=G-v G^3-v^2(G_0^5+G G_0^4)-...\\
...\\
\Rightarrow G_0=G-v G^3-v^2G^5-...
\end{array}\ee
which is then used in the perturbative expansion of the self-energy
(\ref{sigmaper}) to get the formal definition of the skeleton series $S_{LW}$:
\be\label{Sigmaper}\begin{array}{l}
\Sigma=vG_0+v^2G_0^3+...\\
\Rightarrow \Sigma=v G-v^2 G^3-...\equiv S_{LW}[G].
\end{array}\ee
This simplified notation
hides the fact that what here looks like
an algebraic multiplication is in fact an integral/sum over 
the space/time/spin degrees of freedom of the Green's function, 
and each term corresponds to many terms with different combinations 
of multiplications or integrals over arguments.
In the case of the Hubbard Atom, for instance, each term of the perturbative expansion
of the Green's function containing a certain number of $G_0$'s
would involve sums of the variables $n,n',n'',...$ on which each $G_0$ depends.
This means that Perturbation Theory gives rise to a series of
\emph{non-local} functionals of $G_0$.

\begin{figure}[t]
\includegraphics[width=8cm]{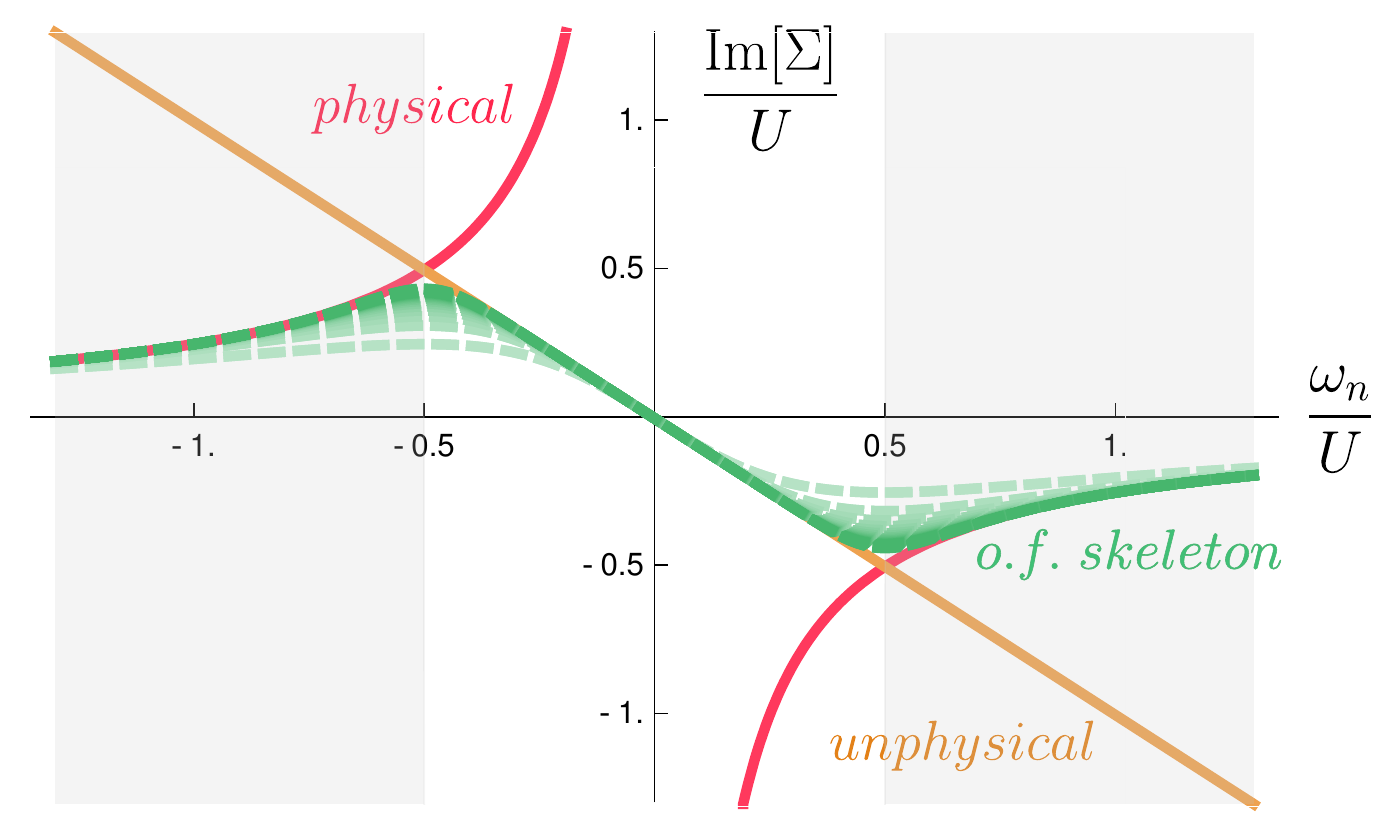}
\caption{%
The physical self-energy $\Sigma(i\omega_n) =-iU^2/(4\omega_n)$ (in red),
the unphysical $\Sigma_u(i\omega_n)=-i\omega_n$ (in orange) and various orders
of the one-frequency--skeleton series $S_{ofs}[\mathcal{G}(i\omega_n)]$
(\ref{SLW}) (in shades of green) are represented. $S_{ofs}[\mathcal{G}(i\omega_n)]$
converges to the physical self-energy only in the region $U^2\leq 4\omega_n^2$ (in gray).
}\label{figPsk}
\end{figure}

However, one can also write down a different, simpler perturbative expansion.
With $\G=F_W[\g]$, expanding $F_W$ in powers of $U$ yields
\be\label{FWtay}
\G=\g+ \g^3 \frac{U^2}{4}+ \g^5\frac{U^4}{16} + ...
\ee
Note that this involves now \textit{actual} multiplications, not convolutions.
The \textit{non-local} character of the general expansion prescribed
from Perturbation Theory is lost, as the above expression
presents a \textit{local} dependence of $\G$ on $\g$;
in other words, at a given value of $z$, $\G$
only depends on $\g$ evaluated at the same $z$.
Obviously, such an expansion is valid only for the Hubbard Atom
and is not as general as Many-Body Perturbation Theory.
Nevertheless, it allows us to define a \emph{local} series, 
by following the steps analogous to equations (\ref{Gper2},\ref{gper},\ref{Sigmaper}).
% ...
This leads to
\begin{multline}\label{SLW}
S_{ofs}[\G]= \frac{U^2 \G}{4}-\frac{U^4 \G^3}{16}+\frac{ U^6 \G^5}{32}+\\
	-\frac{5 U^8 \G^7}{256}+\frac{7 U^{10} \G^9}{512}+...
\end{multline}
which defines the series of functionals
\be
S_{ofs}[f]\equiv \frac{U^2 f}{4}-\frac{U^4 f^3}{16}+\frac{ U^6 f^5}{32}-\frac{5 U^8 f^7}{256}+...
\ee
This formal manipulation of the local series (\ref{FWtay}) allows us to
qualitatively reproduce the results of Kozik \etal for the skeleton series.
In our case, the one-frequency--skeleton series $S_{ofs}$ converges partially to the
correct self-energy, partially to a spurious $\Sigma_u(z)$, as shown in figure
\ref{figPsk}.
As in the case of Kozik \etalcomma the unphysical self-energy corresponds
to the unphysical solution of the inverse problem, $\Sigma_u(z)=\gu^{-1}-\G^{-1}$
(see also figure \ref{figspurious}).
Contrary to what reported in \cite{kozik} about the standard skeleton series,
the convergence of $S_{ofs}$ to the correct result always occurs in a finite
range of $\omega_n$ at fixed $U$:
\be\left\{
\begin{array}{lr}
S_{ofs}[\mathcal{G}(i\omega_n)]=\Sigma(i\omega_n) & \mbox{for }U^2\leq 4\omega_n^2  \\
S_{ofs}[\mathcal{G}(i\omega_n)]=\Sigma_u(i\omega_n) & \mbox{for }U^2\geq 4\omega_n^2.  
\end{array}
\right.\ee
because of the local character of $S_{ofs}$.

Finally, it is important to emphasize that a convergence of the series to an
unphysical self-energy
does not necessarily imply that the corresponding SCDE $X=\g+\g S_{ofs}[X]X$ has,
in the same region, a spurious solution.
In fact, the Dyson equation with a wrong self-energy functional 
may have \textit{no solutions at all}.\footnote{%
	Not enough information was provided in 
	\cite{kozik} to conclude what would happen for the standard
	skeleton series.}
This is indeed our case, where, similar to \cite{stan,rossi,schafer2016}, 
the series that defines $S_{ofs}$ in (\ref{SLW}) converges to 
\be\label{closedSk}
\check{S}[f]\equiv\frac{-1+\sqrt{1+U^2 f^2}}{2 f}
\ee
as long as $|f|\leq 1/U$, which is always the case for physical Green's functions.
Therefore, 
\be\left\{
\begin{array}{lr}
\check{S}[\mathcal{G}(i\omega_n)]=\Sigma(i\omega_n) & \mbox{for }U^2\leq 4\omega_n^2  \\
\check{S}[\mathcal{G}(i\omega_n)]=\Sigma_u(i\omega_n) & \mbox{for }U^2\geq 4\omega_n^2 
\end{array}
\right. \ee
where the functional $\check{S}$ is defined over a larger domain than $S_{ofs}$.
We can therefore study the Dyson equation:
\be\label{dyscheck}
X=\mathcal{G}_0(i\omega_n)+\mathcal{G}_0(i\omega_n) \check{S}[X]X,
\ee
which is satisfied by $X=\mathcal{G}(i\omega_n)$ for $U^2\leq 4\omega_n^2 $,
also outside that region.
To do that, we look at the corresponding equation in one complex variable $x$:
\be\label{equiveqn}
x=\frac{1}{i\omega}
        +\frac{1}{i\omega} \left(\frac{-1+\sqrt{1+U^2 x^2}}{2 x}\right)x.
\ee 
with $\omega\in \mathbb{R}$.
If (\ref{equiveqn}) has no solutions in a certain region of the parameter space,
neither will (\ref{dyscheck}).
We recast (\ref{equiveqn}) as
\be
\frac{2 \omega }{U}=-\frac{i \left(\sqrt{\xi ^2+1}+1\right)}{\xi },
\ee
with $\xi\equiv xU$.
Since we are looking at the region defined by
$U^2>4 \omega_n^2$, we can write 
\be
1>\left|\frac{ \sqrt{\xi ^2+1}+1}{\xi }\right|.
\ee
However, the right-hand side is always greater than 1,
therefore this equation, 
and hence the original Dyson equation (\ref{dyscheck}), 
has no solutions, \textit{Q.E.D.}\footnote{%
	Clearly, this result relies on the closed-form 
	character of the functional (\ref{closedSk}) we considered.
	The use of a truncation of the series $S_{ofs}$, which is always done in practice,
	could introduce more spurious solutions; 
	this happens for example in the simple one-point model of \cite{stan} when only
	the lowest order contribution is used.}

\subsection{Local SIN Series}\label{ssLSS}

In the previous section, we manipulated the perturbative expansion of a local
functional
realizing the map $\g\to\G$, in a way one would do in Perturbation Theory to build
the skeleton series. We showed that the corresponding one-frequency--skeleton series
does not always converge to the correct self-energy.
This supports the interpretation that the numerical findings of \cite{kozik} 
do not point to a failure of the self-consistent approach itself, 
but are rather due to the fact that the skeleton series does not converge to the correct functional. 
Moreover, it suggests that the seed of this failure lies in the way the skeleton
series 
is constructed from the perturbative expansion of $G$.

The problem of building a correct self-energy functional demands a solution.
Following Stan and collaborators \cite{stan},
we attempt therefore, for the simple Hamiltonian here considered,
to build a series that \textit{complements} the one-frequency--skeleton $S_{ofs}$.
As outlined in Section \ref{ssITSCARC}, 
a self-energy functional $S$ can be built via $S[f]\equiv L[f]^{-1}-f^{-1}$
if a functional $L$ is given such that $L[G]=G_0$. 
Such a functional can be connected to a functional $F[G_0]=G$
via $L[G]=F^{-1}[G]$. However, $F$ could be non-injective on a domain bigger than $\{G_0\}$
and one has to find the correct inverse over $\{G\}$.
In our case, we start from $F_W$ defined in (\ref{FW}). This functional is not injective:
its inverse is made of two branches,
\be\begin{array}{l}
F^{-1}_1[f]\equiv \frac{2 \left(\sqrt{U^2 f^2+1}-1\right)}{U^2 f}\\
F^{-1}_2[f]\equiv -\frac{2 \left(\sqrt{U^2 f^2+1}+1\right)}{U^2 f}.
\end{array}\ee
Neither $F_1$ nor $F_2$ are in fact a good $L$ functional, for which $L[G]=G_0$.
What one would need is a \emph{mixture} of the two,
for $F^{-1}_1[G]=G_0$ and $F^{-1}_2[G]=G_0$ on two \emph{complementary} subsets of $\{G\}$:
\be\begin{array}{l}
F^{-1}_1[\G]=\g\;\;\;\;\mbox{for}\;U^2\leq 4\omega_n^2\\
F^{-1}_2[\G]=\g\;\;\;\;\mbox{for}\;U^2\geq 4\omega_n^2\\
\end{array}\ee
The local skeleton series $S_{ofs}$ is obtained by expanding 
$(F_1^{-1}[X])^{-1}-X^{-1}$ in U. Instead,
an expansion of $(F_2^{-1}[X])^{-1}-X^{-1}$ leads to what we shall call
the `one-frequency--SIN' series and denote by $S_{ofSIN}$.\footnote{%
	This is a generalization of the complementary series introduced by Stan 
    and collaborators, which in particular lead to their
	\textit{strong-interaction Hartree-Fock} functional (SIN-HF).}
The two series are related by
\be\label{Psin}
S_{ofSIN}[f]= -\frac{1}{f}-S_{ofs}[f].
\ee
which can be used to calculate $S_{ofSIN}$ once $S_{ofs}$ is provided.

\begin{figure}[t]
\includegraphics[width=8cm]{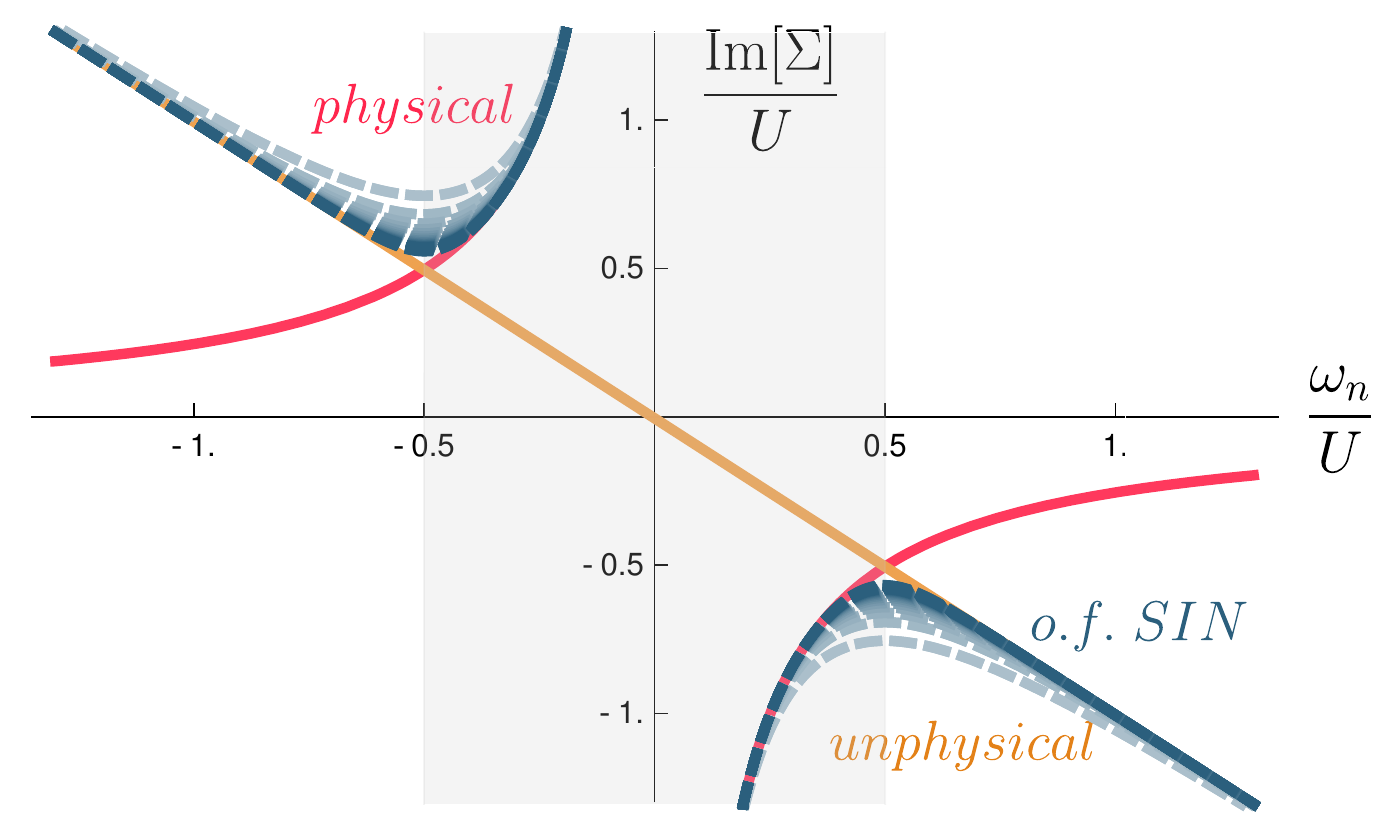}
\caption{The physical self-energy $\Sigma(i\omega_n) =-iU^2/(4\omega_n)$ (in red),
the unphysical $\Sigma_u(i\omega_n)=-i\omega_n$ (in orange) and various orders
of the one-frequency--SIN series $S_{ofSIN}[\mathcal{G}(i\omega_n)]$ (\ref{Psin})
(in shades of blue) are represented.
$S_{ofSIN}[\mathcal{G}(i\omega_n)]$ converges to the physical self-energy only 
in the region $U^2\geq 4\omega_n^2$ (in gray).
}\label{figSIN}
\end{figure}
As shown in fig. \ref{figSIN}, the one-frequency--SIN series evaluated at $\G$
converges to the physical self-energy in the region where $S_{ofs}$ converges to the unphysical one, 
and vice versa.
It follows that the corresponding SCDE $X=\g+\g S_{ofSIN}[X]X$ 
admits $X=\G$ as solution in the range $U^2\geq 4\omega_n^2 $.
An argument similar to the one presented in the previous section allows us
also to state that outside that range the equation has no solutions.
It should be remarked, however, that, even in the pertinent range 
$U^2\geq 4\omega_n^2 $, the actual solution of the self-consistent Dyson equation is not as straightforward as in the 
$U^2\leq 4\omega_n^2 $-range. In particular,
$X=\G$ can not be found by simple
iterative schemes such as $X^{(n+1)}=G_0+G_0 S_{ofSIN}[X^{(n)}]X^{(n+1)}$, and
more sophisticated root-finding algorithms are required.

Finally, we notice that the two functionals $S_{ofs}$ and $S_{ofSIN}$
can be combined to make a proper self-energy functional as
\begin{multline}
\tilde{S}[f]\equiv\theta(zf(z)-1/2)\check{S}[f]+\\
+\left(1-\theta(zf(z)-1/2)\right)\left(-\frac{1}{f(z)}
-\check{S}[f]\right)=\\
=-\frac{1}{f(z)}+\frac{1}{2}\frac{U^2 f(z)}{-1+ {\rm sgn}(zf(z)-\frac{1}{2}) \sqrt{1+U^2 f(z)^2}}
\end{multline}
for which $\tilde{S}[\G]=\g^{-1}-\G^{-1}$ on the entire 
parameter space, like in the case of $S_W$ of (\ref{eq:se}).

\section*{Conclusions}

The self-consistent approach, i.e. the idea of calculating
the one-body Green's function for a generic Hamiltonian problem
via the self-consistent Dyson equation (SCDE) (\ref{scde}), has motivated in the past a notable
amount of research culminating in state-of-the-art methods
for nuclear and condensed matter physics.
However, while the usefulness of the approach and the ``in-principle-exactness'' of the diagrammatic series are most often taken for granted,
a critical analysis and better understanding are still needed. This is also demonstrated by
a recent numerical study by Kozik and collaborators \cite{kozik} that indicates
a failure of the only explicit implementation of the approach at our disposal,
the Luttinger-Ward approach.

In the present work we discussed some general features of the self-consistent approach, and we performed analytical calculations
for 
a specific Hamiltonian problem, the Hubbard atom.
This model is among those used by Kozik \etal 
to test the Luttinger-Ward approach, pointing to failures in some
%
regions of the parameter space. Our calculations allowed us to analyze these failures, explain their origin,
and give indications on how the problems might be overcome.

The first question we addressed was about the possibility to
use the self-consistent approach for this model,
despite the failure of the LW approach found by Kozik and collaborators.
We proved that this can indeed be done.
First we showed that in general the approach can be used,
at least in principle, whenever there is a one-to-one correspondence between
physical interacting and physical non-interacting Green's functions;
then we proved that such a correspondence holds indeed in the case under examination.
Since the set of (non-)interacting physical \GF s does not cover all possible
functions of two arguments (not for the Hubbard atom, nor for the
electronic many-body problem), 
a distinction between the map and the functional that realizes it
becomes essential. The same map $G_0 \to G$ (or $G \to G_0$) can indeed 
be realized by functionals that take different values when evaluated 
\emph{outside} the set of physical Green's functions $\{G_0\}$ (or $\{G\}$).
The distinction between map and functional
also clarifies the apparent contradiction with the conclusion
of Kozik \etal that ``[...] the map $G_0 \to G$ [is] not invertible'',
which, in light of our discussion, 
has to be read as ``the functional used to realize
the map $G_0 \to G$ becomes non-injective 
when evaluated on a domain larger than that of physical $G_0$'s''.

Specifically for the Hubbard atom, 
we have constructed explicit functionals $F_W$ and $L_W$ that realize the maps $G_0 \to G$ and $G \to G_0$,
respectively, on the physical domains of the model. 
These are different from the usual functionals obtained from Perturbation Theory, 
and they have the very important property that all calculations could be done analytically. 
This has, among others, the advantage that one can distinguish problems of convergence from problems of principle.
In particular, these functionals were used to give an explicit realization of the self-consistent approach
(calculating $G$ as solution to $X=G_0+G_0S[X]X$) and the inverse problem 
(calculating $G_0$ as solution of the equation $G=F[X]$).

Having established that, when provided with some functionals for the maps $G_0 \to G$ and $G \to G_0$, 
from which one can build the self-energy functional $S$ to use in the SCDE,
the next question to be addressed was whether the corresponding SCDE $X=G_0+G_0 S[X] X$ has spurious, unphysical solutions alongside the physical one. 
For the general case, we showed that the SCDE has exactly one physical solution
as long as $\{G_0\}\leftrightarrow\{G\}$ holds.
However, since the functional $S$ can take as input functions outside
the set $\{G\}$,  spurious solutions can be avoided if either one restricts
the search over the domain of physical solutions or
makes sure that the functional $L$, linked to $S$ via
$S[f]=L[f]^{-1}-f^{-1}$, is injective over its entire domain.
It should be noted that the conditions to be imposed on the search of $G$ in order to restrict it to
the physical ones are not known in the general case, which means that one can avoid spurious solutions in principle, but the problems may persist in practice.
For the Hubbard atom the conditions restricting $G$ can be formulated in a simple way, which allowed us to illustrate the uniqueness of the solution, 
as well as the fact that spurious solutions appear when the restriction is dropped. 

Once we proved that the self-consistent approach can indeed be used to 
unambiguously determine $G$,
we moved to investigate the failure of the LW approach found by Kozik and collaborators.
In practical applications, one usually takes for the self-energy functional $S[X]$ some approximation to the Luttinger Ward skeleton series which has been obtained from 
rearranging the terms of perturbation theory. It was commonly supposed that one could approach the correct result by including more and more terms of this series. 
However, a counter-example was given by the numerical results of Kozik \etal 

Having a closed functional $F_W$ realizing the map $\{G_0\}\to\{G\}$
for the Hubbard atom, we expanded $F_W$ in powers of $U$ and derived
a perturbative expansion of the \GF\ that maintains the structure of ordinary many-body perturbation theory, but does not contain frequency integrations.
This allowed us to build a self-energy functional 
given as a series of terms similar to the skeleton series, but local in frequency, and hence named `one-frequency--skeleton' series. 
Like $F_W$, this functional  could again be handled fully analytically.
Using the one-frequency--skeleton series to calculate the self-energy
leads to results that are analogous to those of Ref. \cite{kozik};
in particular, the series, when evaluated at a certain $G$,
converges to the correct self-energy only in a limited range of 
parameter space, and it converges to a wrong result outside that region. 
We could relate this finding to the fact that the functional $F_W$ is not injective, such that its inverse, which should realize the map $G\to G_0$, has two branches.
Both of them are needed to build a proper self-energy functional, since 
they both cover only a part of the map between the physical domains of $G$ and $G_0$. 
The one-frequency--skeleton series 
corresponds to the perturbation expansion of one of the two branches and is therefore bound to fail on the other part of the physical domain.

The complementary branch of the self-energy functional can also be expanded in terms of the interaction. For the Hubbard atom we obtained in this way the 
`one-frequency--SIN'-series. Used in the SCDE, the two one-frequency functionals
both lead to the correct result on their respective subdomain, and they both yield no result at all when 
one tries to solve the SCDE outside those respective subdomains. However, spurious solutions can appear when the series are truncated at some order.
The one-frequency--SIN functional is a straightforward generalization of a complementary self-energy functional derived in the context of
 one-point model (OPM) in \cite{stan}. 
This gives hope that it might be possible to go even further, and find a generalized SIN functional 
that would complement the standard skeleton series for the general problem.
Of course, the one-frequency--SIN was derived from the nonperturbative expression we 
used to realize the map $\{G_0\}\to\{G\}$, which for the general problem is not
available.
On the other hand, the simple relation linking the one-frequency--SIN to the
one-frequency--skeleton (\ref{Psin}), 
which has survived the passage from the  OPM
to the Hubbard atom, may indicate the possibility of building such a hypothetical
generalized SIN starting from the skeleton series only, 
or anyway using the tools of Perturbation Theory.

This and many more interesting questions demand
further investigation, in particular: 
For which physical systems will 
the skeleton series lead to serious problems?
Would the corresponding SCDE always be characterized by an unphysical solution that we
can easily recognize as such?
How does the picture change when, as it will be the case in practice, truncations of the series are considered?
How will it be possible to restrict the domain of \GF s in practice? And alternatively, is it worthwhile to maintain the self-consistent approach
in view of its difficulties, or should one rather build self-energy functionals of the non-interacting \GF ?

\acknowledgments{The authors would like to thank all attendees to 
  the Workshop ``“Multiple Solutions in Many-Body Theories''
held in Paris the $13^{th}$ and $14^{th}$ 
of June 2016 for interesting discussions.
The research leading to these results has received
funding from the European Research Council under the European Union’s Seventh Framework Programme
(FP/2007–2013) / ERC Grant Agreement n 320971.
}

\bibliography{bibliography}

\end{document}